\documentclass[aps,twocolumn,prd,preprintnumbers,amsmath,amssymb,superscriptaddress]{revtex4-1}
\pdfoutput=1

\usepackage{graphicx}

\usepackage{hyperref}
\usepackage{mathtools}
\usepackage{amssymb, amsmath, amsopn, amsthm}
\usepackage{epsfig,amsfonts,bm}
\usepackage{graphics}
\usepackage{comment}
\usepackage{color}

\allowdisplaybreaks[1]

\newcommand{\correctedA}[1]{{#1}} 

\newcommand{\fgref}[1]{{\figurename\,\ref{#1}}}

\def\l{\left}
\def\r{\right}
\newcommand{\nl}{\nonumber\\*}

\newcommand{\abs}[1]{{\left|#1\right|}}

\newcommand{\ti}[1]{\tilde{#1}}
\newcommand{\f}[2]{{\frac{#1}{#2}}}
\newcommand{\s}[1]{\sqrt{#1}}

\newcommand{\ket}[1]{\left|#1\right>}
\newcommand{\bra}[1]{\left<#1\right|}

\begin{document}

\title{
Entanglement Entropy of de Sitter Space $\alpha$-Vacua}
\author{Norihiro Iizuka}
\email{iizuka@phys.sci.osaka-u.ac.jp}
\affiliation{
Department of Physics, Osaka University, Toyonaka, Osaka 560-0043, JAPAN
} 

\author{Toshifumi Noumi}
\email{toshifumi.noumi@riken.jp}
\author{Noriaki Ogawa}
\email{noriaki@riken.jp}
\affiliation{RIKEN Nishina Center, Wako 351-0198, JAPAN}


\begin{abstract}
We generalize the analysis of \cite{Maldacena:2012xp}
to de Sitter space $\alpha$-vacua
and compute the entanglement entropy 
of a free scalar
for the half-sphere at late time.
\end{abstract}

\preprint{}


\maketitle



\section{Introduction}\label{sec:intro}
De Sitter space is a very interesting space-time.
It is a solution of Einstein equation 
when cosmological constant dominates,
and it is related to the inflationary stage of 
our universe and also current stage of accelerating universe. 
One peculiar properties of de Sitter space is that 
de Sitter invariant vacuum is not unique; it has a one-parameter family of
invariant vacuum states $\ket{\alpha}$, called $\alpha$-vacua. 

\medskip
\correctedA{The} $\alpha$-vacua give very peculiar behavior for the two point functions in de Sitter space; 
The two point functions on $\alpha$-vacua
between point $x$ and $y$
contain not only the usual short distance singularity $\delta (\abs{x - y})$, where 
$\abs{x-y}$ is de Sitter invariant distances between $x$ and $y$,  but also contain  
very strange singularity such as $\delta (\abs{x - \bar{y}})$ and $\delta (\abs{\bar{x} - y})$, 
where $\bar{x}, \bar{y}$ represent the antipodal points of $x$, $y$. 
Since antipodal points in de Sitter space are not physically accessible due to the separation 
by a horizon, one cannot have an immediate reason to discard two point functions containing 
such an antipodal singularity (See \cite{Spradlin:2001pw} for a nice review, and also \cite{Mottola:1984ar,Allen:1985ux}).
\correctedA{It is therefore unclear
which vacuum should be realized in our universe.
As a result,
a lot of studies have been done
on phenomenological aspects
of the $\alpha$-vacua
(e.g. primordial perturbations
generated during inflation).}

\medskip

Since which vacuum one should choose
is always a very important question,
\correctedA{one is motivated}
to calculate physical quantities 
not only in a particular vacuum but also in others, and 
see if there is a deep reason to choose or discard a particular vacuum. 
\correctedA{In this letter
we compute the entanglement entropy
in de Sitter $\alpha$-vacua.
By generalizing the recent calculation by
Maldacena and Pimentel \cite{Maldacena:2012xp}
in the Euclidean (or Bunch-Davies) vacuum for free scalar fields,
we discuss how entanglement entropy depends on $\alpha$.}

%
%
%
%
%

\section{$\alpha$-vacua of de-Sitter space}

We first introduce the $\alpha$-vacua of de Sitter space in this section.
Let us consider a free real scalar field $\Phi$
of the effective square-mass $m^2$
on de Sitter space
\begin{align}
  I = -\f{1}{2}\int d^4x\sqrt{-g}\l(\partial_\mu\Phi\partial^\mu\Phi + m^2\Phi^2\r)\,.
\end{align}
If we expand the scalar field $\Phi(x)$
in terms of the Euclidean vacuum mode function $\phi_n(x)$ as
\begin{align}
  \label{eq:expand_Phi0}
  \Phi(x) = \sum_n \l(\phi_n(x)a_n + \phi_n^*(x)a_n^\dagger\r)\,,
\end{align}
the Euclidean vacuum $\ket{0}$ is defined by a state satisfying
\begin{align}
\label{Euclidean}
  a_n\ket{0}=0\,.
\end{align}
Here $*$ represents the complex conjugate
and $\dagger$ is the Hermitian conjugate.
The operators $a_n^\dagger$ and $a_n$
are the creation and annihilation operators
on the Euclidean vacuum, respectively.

\medskip
In analogy with (\ref{Euclidean}),
we can introduce a class of states
annihilated by linear combinations of
$a$ and $a^\dagger$
\begin{align}
\label{eq:alpha_oscillators}
  \tilde{a}_n &= (\cosh\alpha) a_n - e^{-i\beta}(\sinh\alpha) a_n^\dagger\,,
\end{align}
where the real parameters $\alpha$ and $\beta$
do not depend on the label $n$ of frequency modes.
In terms of the operator (\ref{eq:alpha_oscillators}),
we introduce a two-parameter family of states defined by
\begin{align}
  \label{eq:alphavac}
  \ti{a}_{n}\ket{\alpha,\beta}= 0\,.
\end{align}
This class of states
are called the $\alpha$-vacua
and it is known that
they reproduce de Sitter invariant Green functions.

\section{Entanglement Entropy on $\alpha$-Vacua}

In this section
we discuss entanglement on the $\alpha$-vacua
of de Sitter spacetime.
Using the same setup and methodology
as the Euclidean vacuum case~\cite{Maldacena:2012xp},
we investigate entanglement at the future infinity.
After clarifying our setup,
we evaluate the density matrix and the entanglement entropy
on the $\alpha$-vacua of free real scalar fields.

\subsection{Setup}
\begin{figure}
  \centering
  \includegraphics[height=5cm, bb=0 0 261 214]{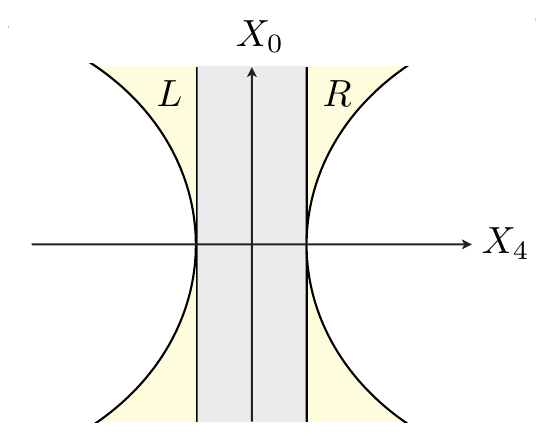}
  \caption{Projection of de Sitter space onto the $(X_0,X_4)$-plane.
We colored $L$ and $R$ with yellow
and $C$ with gray.
Each point represents an $S^2$ with a radius $\s{1+X_0^2-X_4^2}$.}
  \label{fig:deSitter_regions}
\end{figure}
The $4$-dimensional de Sitter space of radius $1$
is defined by a hyperboloid embedded in the $5$-dimensional Minkowski space as
\begin{align}
\label{eq:deSitter}
-X_0^2 + X_1^2 + \dots + X_4^2 = 1
\end{align}
with the Minkowski metric
\begin{align}
ds^2 = -dX_0^2 + dX_1^2 + \dots + dX_4^2\,.
\end{align}
As depicted in \fgref{fig:deSitter_regions},
its projection onto the $(X_0,X_4)$-plane
is given by a region surrounded by the hyperbolae $-X_0^2+X_4^2=1$
because
\begin{align}
  -X_0^2 + X_4^2 = 1- (X_1^2+X_2^2+X_3^2) \le 1\,.
\end{align}
To investigate entanglement
at the future infinity $X_0\to\infty$,
it is convenient to divide
the constant $X_0$ surfaces into
three regions $L$, $C$, and $R$ as~\cite{Maldacena:2012xp}
\begin{align}
L:X_4<-1\,,
\quad
C:-1<X_4<1\,,
\quad
R:X_4>1\,.
\end{align}
Since $L$ and $R$ grow up
as the Minkowski time $X_0$ increases
and $C$ keeps a finite size,
the constant $X_0$ surface
is mostly covered by $L$ and $R$
at the future infinity.
In the following,
we investigate entanglement
of the two regions $L$ and $R$ at the future infinity on the $\alpha$-vacua.

\subsection{Density matrix}

We then discuss entanglement between $L$ and $R$
on the $\alpha$-vacua.
For this purpose,
let us introduce the oscillators
$b_L$ and $b_R$ in the regions
$L$ and $R$,
which satisfy
\begin{align}
\label{eq:b_commutators}
  [\,b_i\,,\,b_j^\dagger\,] = \delta_{ij}\,,
\quad
  [\,b_i\,,\,b_j\,] = [\,b_i^\dagger\,,\,b_j^\dagger\,] = 0\,.
\end{align}
Here and in what follows we drop the label of frequency modes
because different frequency modes are decoupled in the free theory.
The relation between the mode functions in the total space and the subspaces $L$ and $R$ is well studied in \cite{Sasaki:1994yt}.
By using it,
the Bogoliubov coefficients
relating
the annihilation operators $a_{\pm}$ ($\sigma=\pm 1$)
on the Euclidean vacuum in the total space to $b_q$
were evaluated in~\cite{Maldacena:2012xp} as
\begin{align}
\label{eq:abyb}
  a_{\sigma} = \sum_{q=L,R} \left(\gamma_{q\sigma} b_q + \delta^*_{q\sigma}b_q^{\dagger}\right)\,,
\end{align}
where the matrices $\gamma_{q\sigma}$ and $\delta_{q\sigma}$ are
given by
\begin{align}
  \gamma &=
\f{e^{2 \pi  p+i \pi  \nu } (\coth (p \pi )-1) \Gamma \left(i p+\nu +\frac{1}{2}\right)}{4}
\nl
&\qquad\times
\begin{pmatrix}
 \frac{1}{-i+e^{\pi  (p+i \nu )}} & \frac{1}{i+e^{\pi  (p+i \nu )}} \\
 \frac{1}{-i+e^{\pi  (p+i \nu )}} & \frac{-1}{i+e^{\pi  (p+i \nu )}} \\
\end{pmatrix}\,,\\
  \delta &=
\f{\Gamma\left(i p+\nu +\frac{1}{2}\right)}{4\sinh(\pi p)}
\begin{pmatrix}
 \frac{1}{1 + i e^{\pi  (p+i \nu )}} & \frac{1}{1-ie^{\pi  (p+i \nu )}} \\
 \frac{1}{1 + i e^{\pi  (p+i \nu )}} & \frac{-1}{1-i e^{\pi  (p+i \nu )}} \\
\end{pmatrix}\,.
\end{align}
Here $p$ is the Casimir on $H^3$ and $\nu=\sqrt{\f{9}{4}-m^2}$.

\medskip
To discuss entanglement on the $\alpha$-vacua,
we would like to express the $\alpha$-vacua
in the language of the subspaces $L$ and $R$.
By substituting \eqref{eq:abyb} into the definition of the $\alpha$-vacuum oscillators
\eqref{eq:alpha_oscillators},
we obtain
\begin{align}
  \label{eq:ta-b}
  \ti{a}_\sigma = \sum_{q = L,R} \left(\ti{\gamma}_{q\sigma} b_q + \ti{\delta}^*_{q\sigma}b_q^{\dagger}\right)\,,
\end{align}
where
\begin{align}
  \ti{\gamma}_{q\sigma} &= (\cosh\alpha)\gamma_{q\sigma} - e^{-i\beta}(\sinh\alpha)\delta_{q\sigma}\,,\\
  \ti{\delta}_{q\sigma} &= (\cosh\alpha)\delta_{q\sigma}
- e^{i\beta}(\sinh\alpha)\gamma_{q\sigma} 
\,.
\end{align}
For the $\alpha$-vacuum $\ket{\alpha,\beta}$, we adopt an ansatz
\begin{align}
  \label{eq:alphavac_ansatz}
  \ket{\alpha,\beta} = \exp\l(\f{1}{2}\ti{m}_{ij}b_i^\dagger b_j^\dagger\r)\ket{0}_L\ket{0}_R\,,
\end{align}
where $i,j$ run over $\{L,R\}$.
Putting \eqref{eq:ta-b} and \eqref{eq:alphavac_ansatz} into 
the $\alpha$-vacuum condition \eqref{eq:alphavac} and using
the commutation relations \eqref{eq:b_commutators},
 we obtain the condition
\begin{align}
  \ti{m}_{ij} 
= -\ti{\delta}^*_{i\sigma}(\ti{\gamma}^{-1})_{\sigma j}\,.
\end{align}
Furthermore, we introduce a new set of oscillators $\ti{c}_i$ 
in $L$ and $R$ as
\begin{align}
  \label{eq:tc}
  \ti{c}_i = u_i b_i + v_i b_i^\dagger
\quad
  \l( \abs{u_i}^2 - \abs{v_i}^2 = 1 
\r)
\end{align}
such that the wavefunction of the $\alpha$-vacuum is diagonalized as
\begin{align}
\label{eq:alphavacbyc}
  \ket{\alpha,\beta} = \exp(\ti\kappa \ti{c}_L^\dagger\ti{c}_R^\dagger)\ket{\ti{0}'}_L\ket{\ti{0}'}_R
\l(= \sum_{n\ge 0}\ti\kappa^n\ket{\ti{n}'}_L\ket{\ti{n}'}_R\r)\,,
\end{align}
where $\ket{\ti{0}'}_i$ is the ``vacuum'' for $\ti{c}_i$,
i.e.,
\begin{align}
  \ti{c}_i\ket{\ti{0}'} = 0\,.
\end{align}
For the normalizability of $\ket{\alpha,\beta}$, 
we need $\abs{\ti\kappa}<1$.
Furthermore, from \eqref{eq:alphavacbyc} we find
\begin{align}
  \ti{c}_L\ket{\alpha,\beta} = \ti\kappa\ti{c}_R^\dagger\ket{\alpha,\beta}
\quad\text{and}\quad
  \ti{c}_L\ket{\alpha,\beta} = \ti\kappa\ti{c}_R^\dagger\ket{\alpha,\beta}\,.
\end{align}
\begin{widetext}
By using \eqref{eq:alphavac_ansatz} and \eqref{eq:tc},
this condition can be rewritten in terms of the $b_i$ oscillators
and finally results in
\begin{align}
  \begin{pmatrix}
    \ti\rho & 1 & 0 & -\ti\zeta\ti\kappa \\
    \ti\zeta & 0 & -\ti\kappa & -\ti\rho\ti\kappa \\
    0 & -\ti\zeta^*\ti\kappa^* & \ti\rho^* & 1 \\
    -\ti\kappa^* & -\ti\rho^*\ti\kappa^* & \ti\zeta^* & 0
  \end{pmatrix}
  \begin{pmatrix}
    u_R \\ v_R \\ u_L^* \\ v_L^*
  \end{pmatrix}
  = 0\,,
\end{align}
where 
\begin{align}
\label{eq:tilderho}
\ti{\rho}&\equiv\ti{m}_{LL}=\ti{m}_{RR}\nl
&= 
\frac{  
-\left(1+e^{2 i \pi  \nu }\right)e^{2\pi p} \left(\sinh^2\alpha + e^{2 i (\beta +\pi  \nu )}\cosh^2\alpha \right)+\left(e^{2 \pi  p}-1\right)^2 e^{i (\beta +2 \pi  \nu )}\sinh\alpha \cosh\alpha}{\left(e^{2 \pi  p}+e^{2 i \pi  \nu }\right) \left(\sinh^2\alpha + e^{2 \pi  p+2 i (\beta +\pi  \nu )}\cosh^2\alpha\right)}\,,
\\
\label{eq:tildezeta}
\ti{\zeta}&\equiv\ti{m}_{LR}=\ti{m}_{RL}\nl
&=
-\frac{i \left(e^{2 \pi  p}-1\right) e^{\pi  (p+i \nu )} \left(\sinh\alpha+e^{i \beta } \cosh\alpha\right) \left(\sinh\alpha +e^{i (\beta +2 \pi  \nu )}\cosh\alpha \right)}{\left(e^{2 \pi  p}+e^{2 i \pi  \nu }\right) \left(\sinh^2\alpha + e^{2 \pi  p+2 i (\beta +\pi  \nu )}\cosh^2\alpha \right)}\,.
\end{align}
In order that this linear equation has nontrivial solutions, the determinant of this $4\times 4$ matrix has to vanish. It leads to a simple equation
\begin{align}
\label{eq:tildekappa_eq}
  \abs{\kappa}^4
  - 2\ti{\Lambda} \abs{\kappa}^2
  + 1 = 0
  \quad
  {\rm with}
  \quad
  \ti{\Lambda} = \f{\big|\ti\zeta\big|^4 + (\abs{\ti\rho}^2 -1)^2 - (\ti\rho^2\ti\zeta^{*2} + \ti\rho^{*2}\ti\zeta^2)}{2\big|\ti\zeta\big|^2}\,.
\end{align}
We then obtain 
\begin{align}
\label{eq:tildekappa}
  \abs{\ti\kappa}^2 = \ti\Lambda - \sqrt{\ti\Lambda^2 -1 }\,,
\end{align}
where we chose a solution satisfying the normalizability condition $|\kappa|<1$.
From \eqref{eq:alphavacbyc},
the normalized reduced density matrix $\ti\rho_L$ is computed as
\begin{align}
  \ti\rho_L = \f{1}{1-\abs{\ti\kappa}^2}\sum_{n=0}^{\infty}\abs{\ti\kappa}^{2n}\ket{\ti{n}'}_L\bra{\ti{n}'}_L\,.
\end{align}
It should be noticed that
the density matrix $\ti\rho_L$ is invariant
under the shift $\nu\to\nu+1$
because $\ti\rho$ and $\ti\zeta$ are invariant
up to an overall sign factor,
and \correctedA{so}
$\ti\Lambda$ and $\ti\kappa$ are invariant under the shit.
\correctedA{
When $\beta=0$,
it is also invariant under a reflection at $\nu=1/2$ or $\nu=1$,
i.e. $\nu\to 1/2-\nu$ and $\nu\to 1-\nu$.
}


\end{widetext}

\subsection{Entanglement entropy}

\begin{figure}
  \centering
  \includegraphics[width=0.4\textwidth]{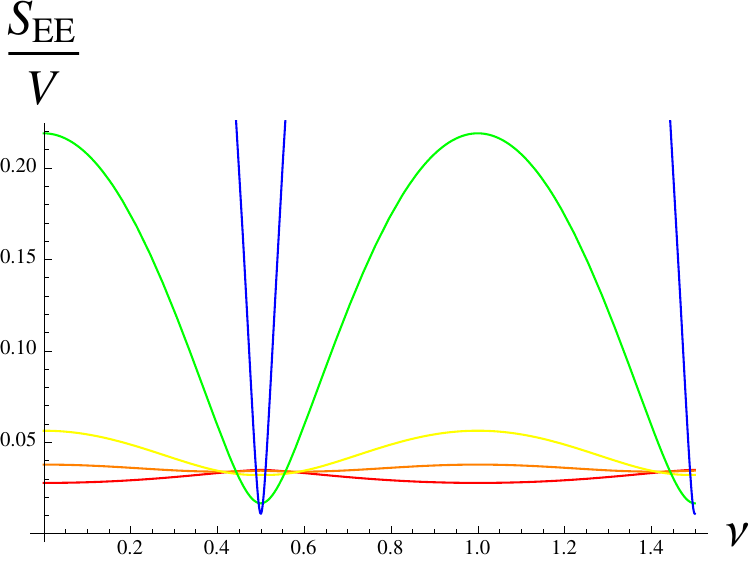}
  \caption{Plot of $S_{EE}/V$ against $\nu$, for $\alpha=0$ (red), $0.1$ (orange), $0.25$ (yellow), $1$ (green), $2$ (blue).
Notice the periodicity 
\correctedA{
and reflection symmetries.
}
}
  \label{fig:SEE_nuplot}
\end{figure}

\begin{figure}
  \centering
  \includegraphics[width=0.4\textwidth]{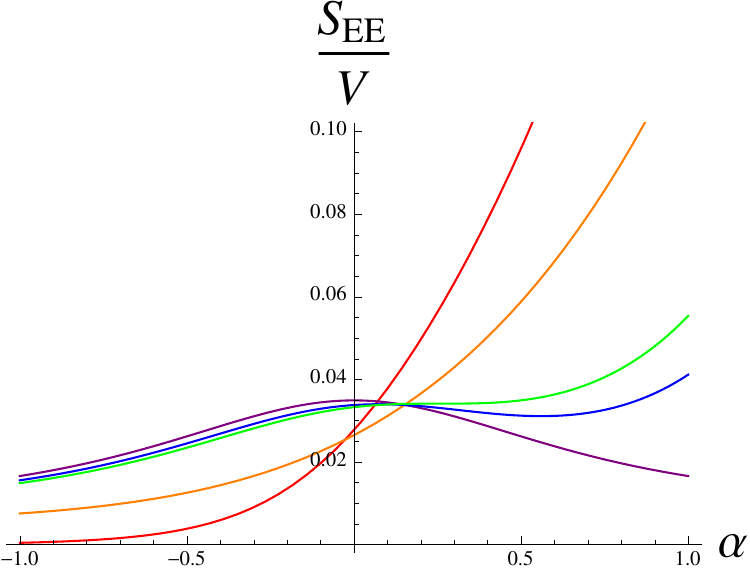}
  \caption{Plot of $S_{EE}/V$ against $\alpha$, 
\correctedA{for $\nu=0$ (red), $0.25$ (orange), $\nu_c=0.4062\dots$ (green), $0.43$ (blue) and $0.5$ (purple).}}
  \label{fig:SEE_alphaplot}
\end{figure}

Finally, let us evaluate the entanglement entropy
using the obtained density matrix $\ti\rho_L$.
The entanglement entropy for each frequency mode
is
\begin{align}
\nonumber
S_{EE}(p) 
&= - \text{Tr}\rho_L(p)\log \rho_L(p)\\
&= -\log(1-\abs{\ti\kappa}^2) - \f{\abs{\ti\kappa}^2}{1-\abs{\ti\kappa}^2}\log\abs{\ti\kappa}^2\,,
\end{align}
where note that $\ti{\kappa}$ has a $p$-dependence.
The total entanglement entropy per volume is therefore given by
\begin{align}
\label{eq:SEEperV}
  S_{EE}/V = \int_0^{\infty}\!dp\, \mathcal{D}(p)S_{EE}(p)\,,
\end{align}
where $V$ is the spatial volume of the $L$ region ($\simeq H^3$)
and the state density $\mathcal{D}(p)$ is
\begin{align}
  \mathcal{D}(p) = \f{p^2}{2\pi^2}\,.
\end{align}
This EE density \eqref{eq:SEEperV} is equal to the logarithmic coefficient
$2\pi S_{\mathrm{intr}}$,
in the terminology of \cite{Maldacena:2012xp}.
Using these formulas,
we numerically plotted the entanglement entropy
in \fgref{fig:SEE_nuplot} and \fgref{fig:SEE_alphaplot}.

\correctedA{
In \fgref{fig:SEE_nuplot},
the periodicity and reflection symmetries are manifest.
\fgref{fig:SEE_alphaplot} shows that the entanglement entropy
blows up for positively large $\alpha$ in $0\le\nu<1/2$.
It monotonically increases against $\alpha$ for small $\nu$,
and a local minimum point appears when $\nu$ reaches $\nu_c=0.4062\dots$,
at $\alpha_m(\nu_c)=0.2576\dots$.
After that $\alpha_m(\nu)$ increases as $\nu$
and disappears to infinity at $\nu=1/2$.
}

\section{Discussion}

In this \correctedA{letter},
we have shown the calculation of the entanglement entropy for 
de Sitter space $\alpha$-vacua, 
by generalizing the analysis of \cite{Maldacena:2012xp}. 
As is seen in \fgref{fig:SEE_nuplot} and \fgref{fig:SEE_alphaplot},
entanglement entropy increases significantly 
as we 
\correctedA{
take $\alpha$ very large,
for generic values}
of $\nu$.
However 
\correctedA{only for $\nu = 1/2$ and $3/2$,}
this tendency 
\correctedA{disappears.}
Note that 
\correctedA{$\nu = 1/2$}
is the conformal mass
\correctedA{and $\nu=3/2$ is massless.}
It is interesting to understand more physically why such a mass dependence occurs.

\medskip
Our calculation is done in the free scalar field.
Therefore
direct comparison with the holographic calculation for the Euclidean vacuum~\cite{Maldacena:2012xp} is 
difficult.
It must be interesting to ask how the calculation of entanglement entropy on the $\alpha$-vacua
can be done in the strong coupling limit via holography%
\correctedA{,}
a la Ryu-Takayanagi formula \cite{Ryu:2006bv}.
Understanding these will hopefully shed more light on the question of which vacuum one should 
choose in de Sitter space. 
We hope to come back to these question in near future. 

\medskip
{\bf Note added:}
Even though we have finished the calculation in this 
\correctedA{letter}
long before,
\correctedA{we were working to include a holographic analysis.}
Then, a paper~\cite{Kanno:2014lma} appeared, which overlaps significantly to our work.
Note that the result eq.~(3.16) in \cite{Kanno:2014lma} coincides with our results \eqref{eq:tilderho}-\eqref{eq:tildekappa}.   


\acknowledgements
We are very grateful to Akihiko Ishibashi and Kengo Maeda
for discussions and collaboration in the initial stage of this work.
This work was partially supported by the RIKEN iTHES Project.  
NI is also supported in part by JSPS KAKENHI Grant Number 25800143.
The works of TN and NO are supported by the Special Postdoctoral
Researcher Program at RIKEN. 
TN also thanks
Institute for Advanced Study at the Hong Kong University of Science and Technology,
where a part of this work was done.

\vspace{1cm}

\bibliography{reference}

\end{document}